# Bayesian Optimization in Continuous Spaces via Virtual Process Embeddings

Mani Valleti,[1,a] Rama K. Vasudevan,[2] Maxim A. Ziatdinov,[2,3] and Sergei V. Kalinin[4,b]

[1] Bredesen Center for Interdisciplinary Research, University of Tennessee, Knoxville, TN 37916 USA

[2] Center for Nanophase Materials Sciences, Oak Ridge National Laboratory, Oak Ridge, TN 37831 USA

[3] Computational Sciences and Engineering Division, Oak Ridge National Laboratory, Oak Ridge, TN 37831 USA

[4] Department of Materials Science and Engineering, University of Tennessee, Knoxville, TN 37916 USA

Automated chemical synthesis, materials fabrication, and spectroscopic physical measurements often bring forth the challenge of process trajectory optimization, i.e., discovering the time dependence of temperature, electric field, or pressure that gives rise to optimal properties. Due to the high dimensionality of the corresponding vectors, these problems are not directly amenable to Bayesian Optimization (BO). Here we propose an approach based on the combination of the generative statistical models, specifically variational autoencoders, and Bayesian optimization. Here, the set of potential trajectories is formed based on best practices in the field, domain intuition, or human expertise. The variational autoencoder is used to encode the thus generated trajectories as latent vector, and also allows for the generation of trajectories via sampling from latent space. In this manner, Bayesian Optimization of the process is realized in the latent space of the system, reducing the problem to a low-dimensional one. Here we apply this approach to a ferroelectric lattice model and demonstrate that this approach allows discovering the field trajectories that maximize curl in the system. The analysis of the corresponding polarization and curl distributions allows the relevant physical mechanisms to be decoded.

[a] svalleti@vols.utk.edu
[b] sergei2@utk.edu

Automated experiment and autonomous laboratories are rapidly becoming the leading paradigm in areas ranging from materials science and condensed matter physics to chemistry. For synthesis, the concepts such as fully autonomous robotic laboratories,[1-3] microfluidic systems,[4] and integrated human-high throughput synthesis and characterization workflows[5] have been proposed. In microscopy and scattering, automated experiment for rapid optimization of imaging parameters and discovery is rapidly gaining popularity,[6, 7] the development stimulated by the emergence of the control interfaces by leading manufacturers. Notably, microscopic imaging over the composition spread libraries opens further opportunities for rapid materials discovery.

Common for materials discovery and optimization and imaging problems is the need to explore non-convex parameter spaces towards desired functionality.[8-11] For materials synthesis, this is often the compositional space of selected multicomponent phase diagram,[12, 13] or the processing history.[14, 15] For imaging, this is the image plane of the object[16] or the parameter space of the microscope control.[17] From the optimization perspective, critical considerations are the dimensionality and completeness of the parameter space, and the properties of the target function defined over it.

For compositional space in synthesis or instrumental tuning alike, the parameter space is generally low dimensional and complete, often allowing for the use of the classical Bayesian Optimization based methods.[18-21] In cases where the function defined over the parameter space is discontinuous, the classical BO methods fail. However, if the character of these behaviors is partially known, the use of the structured Gaussian Processing with the mean function tailored for specific behavior allows to address this problem. [22]

However, the many problems including molecular and materials discovery[23], active learning of structure-property relationships,[24, 25] and process optimization[26] require optimization in very large dimensional spaces, e.g., the chemical space of the system,[27] possible processing trajectories, or local microstructures. It is well recognized that solution for this problem is possible if the behaviors of interest form low-dimensional manifolds in these high dimensional spaces, and hence active learning problem is indelibly linked to the discovery of this manifold.

Recently, we have demonstrated the use of the deep kernel learning methods for discovery of structure-property relationships. [28-31] Here, the algorithm has the access to the full structural information in the system and seeks to explore the relationship between the structure and functionalities by sequential measurements. The deep kernel learning parts build the correlative relationship between structure (e.g., image patch) and properties (spectra), whereas the use of physics-based search criteria allows to guide the discovery towards specific objects of interest.

Here, we extend this concept towards optimization in high dimensional spaces for applications such as materials processing or adaptive imaging and spectroscopy in physical measurements. We propose that from a practical viewpoint, we aim to optimize not the full set of possible trajectories, but a practically relevant subset that contains sufficient variability to approach a desired target. These can be chosen based on domain expertise, prior published data, or domain-specific intuition. One way to achieve this will be to use the trajectories parametrized via a certain functional form. However, this approach leads to large parameters spaces as well.

Hence, we introduce the concept of deep kernel learning on the virtual spaces and demonstrate it for the model lattice phase model of ferroelectric material. However, the proposed approach is universal and can be applied to other processing problems.

To illustrate this concept, we first explore the salient features of low-dimensional representations of complex functions. As an example, we consider the set of arbitrary trajectories, each represented by a linear combination of Legendre polynomials,

$$F(x) = \sum_{p=0}^{N} A_p L_p(x) \tag{1}$$

Where $L_p(x)$ is the Legendre polynomial of degree $p$ and $A_p$ is its coefficient in the linear expansion of the function $F(x)$. The Legendre polynomials $L_p$ form an orthogonal basis on the interval [0,1], and possess the properties that $L_p(0) = -1/+1$ for odd/even $p$, and $L_p(1) = 1$. As an orthogonal system over the unit interval, they form a convenient basis to expand any practically relevant function, and the properties of such expansions are well explored.

Here, as an illustration, we generate 5000 trajectories governed by equation. (1), where coefficients $A'_p$ are sampled from a uniform distribution on [0, 1] and $N = 14$. These randomly sampled coefficients are then scaled in the form $A_p = A'_p/p$ to reduce the effects of higher-order polynomials and make them more practical. Twenty-five randomly sampled trajectories from the family generated are shown in Fig. 1a. The family of trajectories generated lie in a 15-dimensional space where $p^{th}$ dimension corresponds to the corresponding Legendre polynomial $L_p$ and by construction, will contain the trajectories that represent practically important processing trajectories, e.g., temperature profile during field annealing, field evolution during the poling of ferroelectric material, etc. However, optimization in such high-dimensional space is complex.

We subsequently explore the encoding of the generated functions into a low dimensional latent space via the variational autoencoder (VAE) described in our previous works. [25, 32-35] Here, the trajectories generated as described above act as the input to the VAE. The VAE then builds the smooth encoding of trajectories, where the trajectories are mapped to an n-dimensional continuous latent space. Each trajectory is then represented as an n-dimensional vector ($l_1, …, l_n$) in the latent space, where $l_n$ is the $n^{th}$ component of the latent vector. The dimensionality of the latent space ($n$) is a user selected variable.

The dimensionality of the latent space is set to 2 for visualization purposes. The two-dimensional latent space is uniformly sampled and each sampled point in the latent space is then decoded back into the space of trajectories. This decoded latent space shown in Fig. 1b, helps in visualizing the trajectories encoded in different parts of the latent space. We will refer to such plots as the decoded latent space from now on. It can be observed from the decoded latent space that the trajectories are encoded in a continuous and smooth manner, which is the characteristic of the VAE. Each trajectory in the real space is represented as a scatter-point in the latent space and then colored using the *rmse* between the original function and the output of VAE. The reconstruction error here is the average of root mean squared error (*rmse*) between the trajectories in the input space and the encoded-decoded functions, the outputs of the VAE. This plot is shown in Fig. 1c, and it can be observed from this figure that the reconstruction error is not a function of latent space, i.e., there is no specific region in the latent space where the trajectories are encoded poorly by VAE. We will refer to such plots where trajectories are plotted as scatter points in the latent space as the latent distributions. Finally, to assess the effect of latent space dimensionality on the encoding, we plot the reconstruction error vs. the number of latent dimensions in Fig. 1d. Initially, the error decreases with the number of latent dimensions and increases slightly. This is because with a large number of latent dimensions, the latent space starts to encode the noise in the functions.

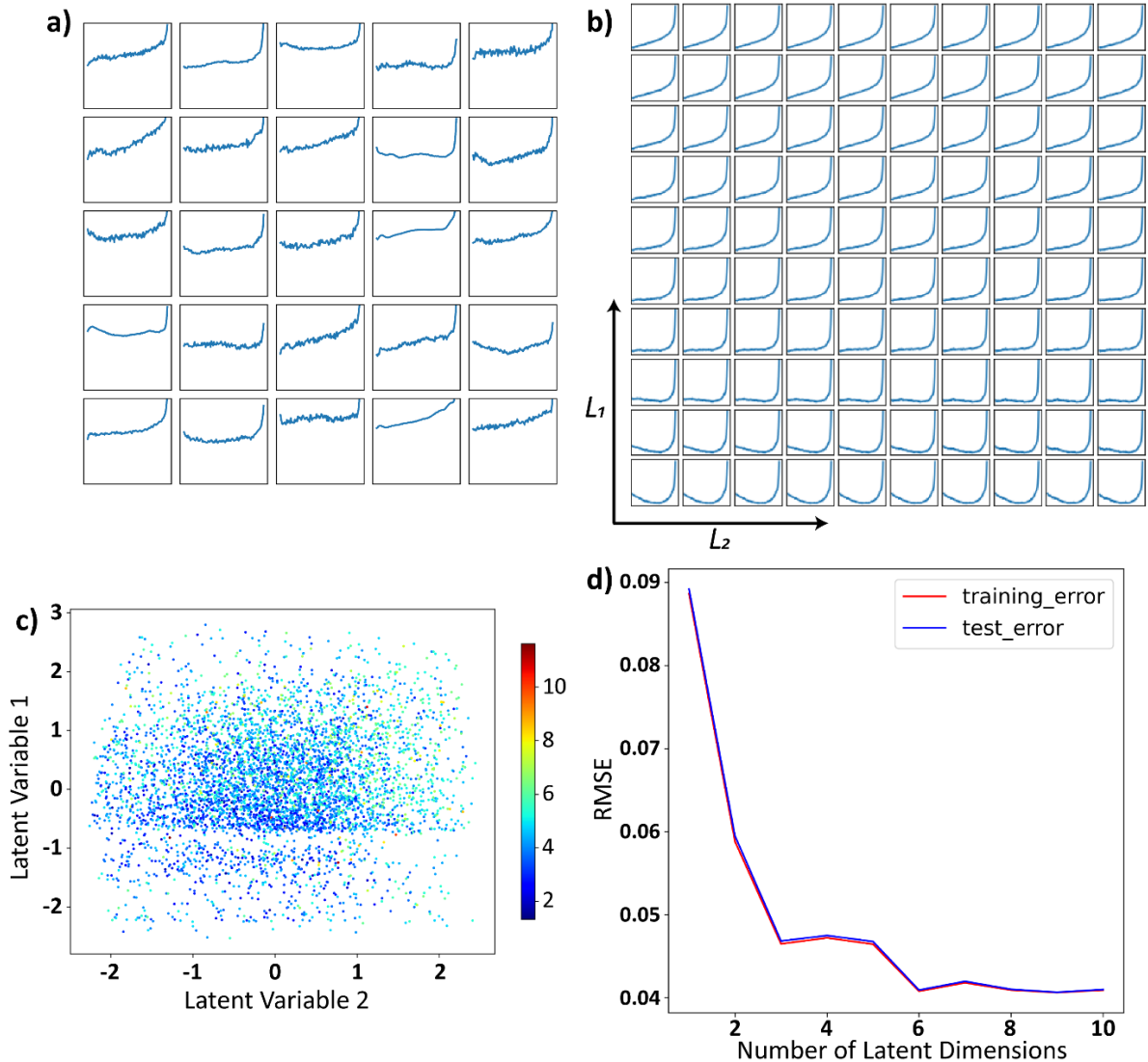

**Figure 1.** (a) Twenty-five randomly selected functions generated by equation. 1, (b) Uniformly sampled and decoded latent space, (c) Functions generated by equation. 1 in the latent space and are colored using RMSE, illustrating the reconstruction quality vs. latent encoding, (d) RMSE

between the function in the input space and the functions that are decoded back to the input space after encoding as a function of number of latent dimensions.

To get insight into the relationship between original parameters that generated the trajectories and the latent distribution, we plotted the latent distribution with colors corresponding to the parameters. Latent distributions where the trajectories are encoded into latent space, are color coded using the $A_1$, $A_2$, and $A_4$ are shown in Fig. 2a-c respectively. The latent space appears to have a strong dependence on $A_1$ (Fig. 2a) and $A_2$ (Fig. 2b) while it does not appear to have any correlation with $A_4$ (Fig. 2c). This is intuitively clear, as the first two coefficients in the linear expansion of the functions have larger weights by construction. The dependence of the latent space on the coefficients of the higher order polynomials diminishes with the order and can be explored further in the attached Jupyter notebook in the data availability statement.

However, the latent space can be sampled to form trajectories in the real space with a continuous variability of higher-order parameters. To demonstrate this property, we uniformly sampled the latent space into 10000 points where each latent dimension varies between [-2, 2]. These latent points are then decoded into the real space and are expanded to a Legendre polynomial series using the least squares method i.e., ($l_1$, $l_2$) => f(x) into Legendre series, $f(x) = \sum_{p=0}^{\infty} \tilde{A}_p L_p(x)$. With this, we plot the expansion coefficients in the latent space as shown in Fig. 2d-e. To show that the latent space has a smooth variation in both lower order and higher order polynomials, we used coefficients $\tilde{A}_1$, $\tilde{A}_2$, $\tilde{A}_3$ in fig. 2d and $\tilde{A}_9$, $\tilde{A}_{10}$, $\tilde{A}_{11}$ in fig. 2e. The value of these coefficients in the expanded Legendre series act as color channels for the RGB images. This would aid is noticing the variability of three coefficients per image. Note that while the resulting distributions are generally multi modal, they nonetheless show clear variability over latent space for lower and higher order coefficients. In such a way, the process trajectories are effectively encoded in the latent space, and corresponding parameters form smooth fields over the latent space with a small number of minimum and maxima. Assuming that the relationship between the trajectory of the process and resultant materials functionality is smooth almost everywhere, this strongly suggests that the Bayesian optimization can be performed over the latent space, where the decoded trajectories are used as an input to the process, and resultant functionality defines the target function.

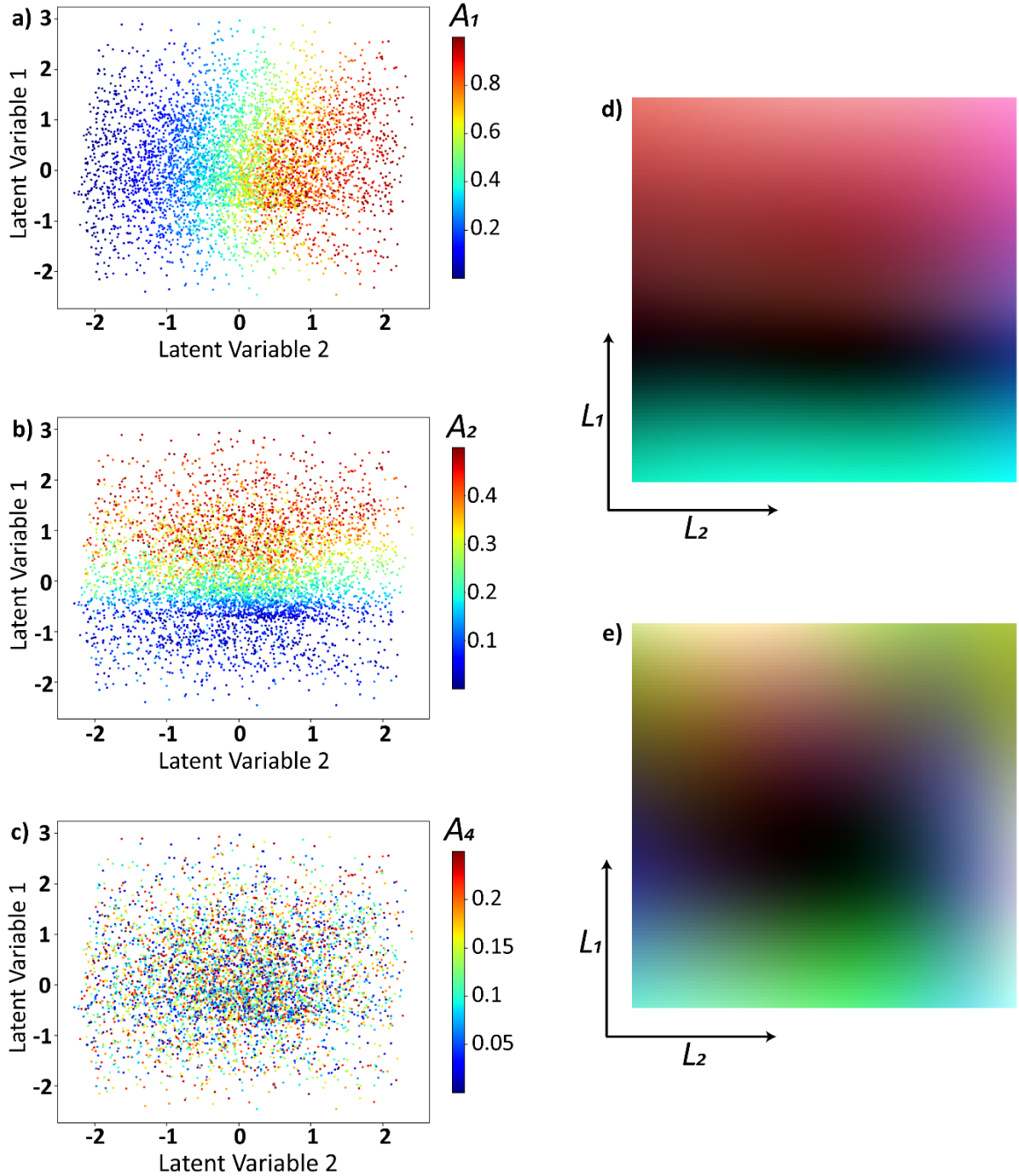

**Figure 2.** Input functions represented as points in the latent space and are colored as a function of (a) $A_1$, (b) $A_2$, (c) $A_4$. Latent space is uniformly sampled and decoded into the space of input function which are then expanded into a Legendre series and the coefficients (d) $\tilde{A}_1$, $\tilde{A}_2$, and $\tilde{A}_3$, (e) $\tilde{A}_9$, $\tilde{A}_{10}$, and $\tilde{A}_{11}$ are plotted in the latent space as color channels in an RGB image.

Prior to the optimizing the latent space with Bayesian Optimization, we assess the effect of the presence of trajectories in the input dataset that are not part of the physical system on VAE. We generated a dataset where a fixed proportion of these trajectories are linear (which comply with the physical system), and the remaining are sinusoidal. The linear trajectories have variability in slope and *y*-intercept, which leads to variability in their range. For sinusoidal trajectories, only their frequency has been varied. In this section, we refer to the sinusoidal trajectories as wrong trajectories, i.e., the trajectories that do not belong to the original distribution.

We trained three different VAEs with varying proportions (1%, 10%, 30%) of wrong trajectories in the input data. The decoded latent space for each of the three different wrong trajectory proportions are shown in Fig3a (1%), Fig. 3b (10%), and Fig 3c (30%). It can be observed from the results that the latent space is affected by the presence of wrong trajectories irrespective of their proportion in the input data. Furthermore, all the input functions are plotted as points in the latent space where the linear trajectories are colored blue, and the sinusoidal trajectories are colored brown. This latent space representation of linear and wrong trajectories is shown in Fig3d (1%), Fig. 3e (10%), and Fig. 3f (30%). In all the three cases, the sinusoidal trajectories occupy a very tiny portion of the latent space. This is because the variability in the linear trajectories is much higher than the sinusoidal trajectories which are only varied in frequency. The Jupyter notebook provided in the data availability statement shows an example in which 90% of the curves in the input curves are sinusoidal and the discussion on latent space still holds true. From this analysis, it can be deduced that the latent space is always affected even if there is a limited presence of wrong trajectories. The families of curves are grouped together in the latent space whose assigned area in the latent space is proportional to their variability.

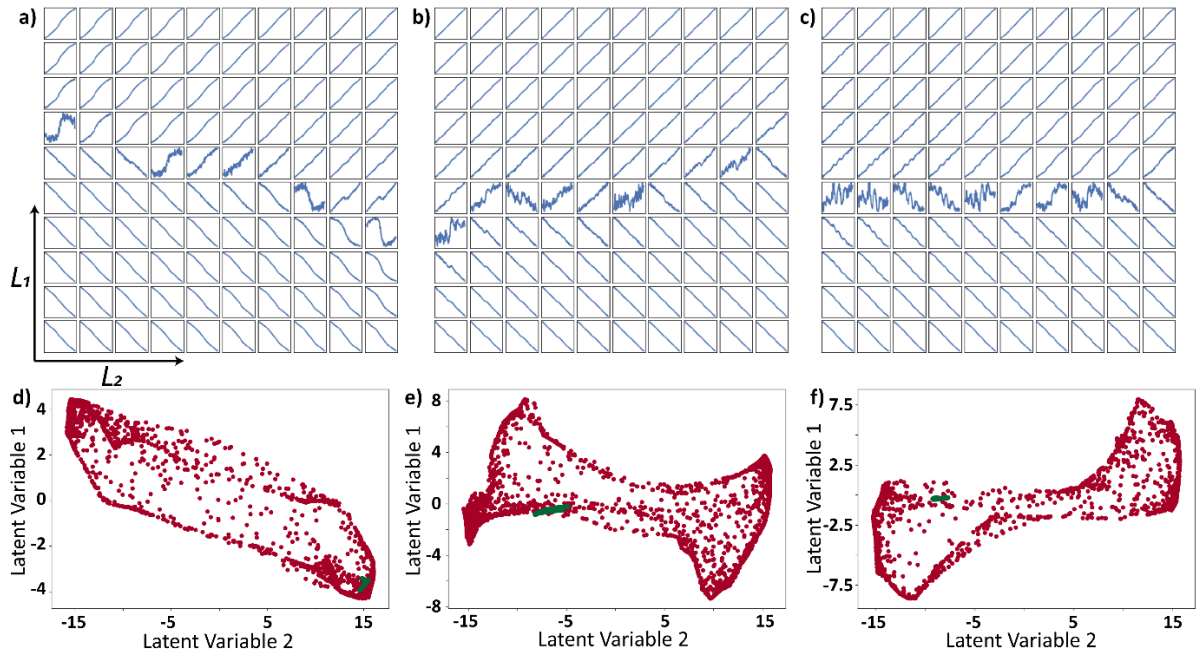

**Figure 3.** Decoded latent space when a) 1%, b) 10%, and c) 30% of the input trajectories (linear) are corrupted with the trajectories that do not follow the same physics of the system. Distribution of input dataset in the latent space for the presence of d) 1%, e) 10%, and f) 30% wrong trajectories

in the input system. The red dots represent the linear (right trajectories) and the green dots represent the sinusoids (wrong trajectories) in the latent space.

With this efficient encoding of the complex multiparametric trajectories in low dimensional parameter space demonstrated, we further extend this approach towards the process optimization. Here, we implement the Bayesian Optimization over pre-encoded set of trajectories. The latter in turn is generated based on domain specific intuition and prior knowledge, or can be derived assuming some desirable factors of variability for a given set of functions. The latent vector is decoded to yield the process trajectory, and the experiment is run using this trajectory. The experimental output is used as a signal to guide the BO over the latent space.

As a model system, we implement the analysis of the domain structure evolution in the ferroelectric material. Here, we use the FerroSIM model, representing the discrete lattice of continuous spins as described in Ref. [36] However, the choice of this specific example does not affect the generality of proposed approach. FerroSIM is a lattice model representing a ferroelectric material introduced by Ricinshi et al. [37] In this model, the order parameter, polarization in the case of a ferroelectric material is represented as a continuous quantity at every lattice site. The local free energy function is assumed to be of the Ginzburg-Landau (GLD) form and is given by equation 2.

$$F_{ij} = \alpha_1 \left( p_{x_{ij}}^2 + p_{y_{ij}}^2 \right) + \alpha_2 \left( p_{x_{ij}}^4 + p_{y_{ij}}^4 \right) + \alpha_3 p_{x_{ij}}^2 p_{y_{ij}}^2 - E_{loc_{x_{ij}}} p_{x_{ij}} - E_{loc_{y_{ij}}} p_{y_{ij}} \quad (2)$$

Where $\alpha_1$, $\alpha_2$, and $\alpha_3$ are the GLD coefficients, $(i, j)$ refers to the row and column index of a lattice site, $p_x$ and $p_y$ are the components of local polarization vector in x and y direction respectively at the lattice site referred by $(i, j)$, $-E_{loc}.p$ terms are the resultant free energy locally due to the coupling between polarization ($p$) and the electric field. $E_{loc}$ is the local field at the lattice site and is given by the equation 3. It is a vector sum of external field ($E_{ext}$), depolarization field ($E_{dep}$), and any random field disorders ($E_d(i, j)$) present at the lattice site. The depolarization field is assumed to be proportional to the total polarization of the lattice and is given by equation 4.

$$E_{loc} = E_{ext} + E_{dep} + E_d(i,j) \quad (3)$$

$$E_d = -\alpha_{dep} <p>, where <p> = \sum_N p \quad (4)$$

The total free energy of the lattice is then a sum of the local free energies and the interactions between the lattice sites and their neighbors. We will only be considering the nearest neighbor interactions for this analysis; complex models can be constructed from this model where the interactions last several unit cells. The total free energy of the lattice system is given by the equation 5, where $K$ is the coupling constant and the interactions are summed over the entire neighborhood via summation over *k,l*. Summation over *i, j* results in total interaction energy over the entire lattice.

$$F = \sum_{i,j}^{N} F_{ij} + K \sum_{k,l} (p_{x_{ij}} - p_{x_{i+k,j+l}})^2 + K \sum_{k,l} (p_{y_{ij}} - p_{y_{i+k,j+l}})^2 \quad (5)$$

Finally, the time dynamics of the system are given by the classical Landau-Khalatnikov equation and is given by equation 6. The polarization at each time step is then updated by calculating the derivatives of the total free energy with respect to the polarization field at the previous time step and are given by equations 7 and 8.

$$\frac{\gamma dp_{i,j}}{dt} = -\frac{\partial F}{\partial p_{i,j}} \quad (6)$$

$$\frac{dp_{x_{ij}}}{dt} = -\gamma^{-1} \left( 2\alpha_1 p_{x_{ij}} + 4\alpha_2 p_{x_{ij}}^3 + 2\alpha_3 p_{x_{ij}}^2 p_{y_{ij}} + 2K \sum_{k,l} \left( p_{x_{ij}} - p_{x_{i+k,j+l}} \right) - E_{loc} \right) \quad (7)$$

$$\frac{dp_{y_{ij}}}{dt} = -\gamma^{-1} \left( 2\alpha_1 p_{y_{ij}} + 4\alpha_2 p_{y_{ij}}^3 + 2\alpha_3 p_{x_{ij}} p_{y_{ij}}^2 + 2K \sum_{k,l} \left( p_{y_{ij}} - p_{y_{i+k,j+l}} \right) - E_{loc} \right) \quad (8)$$

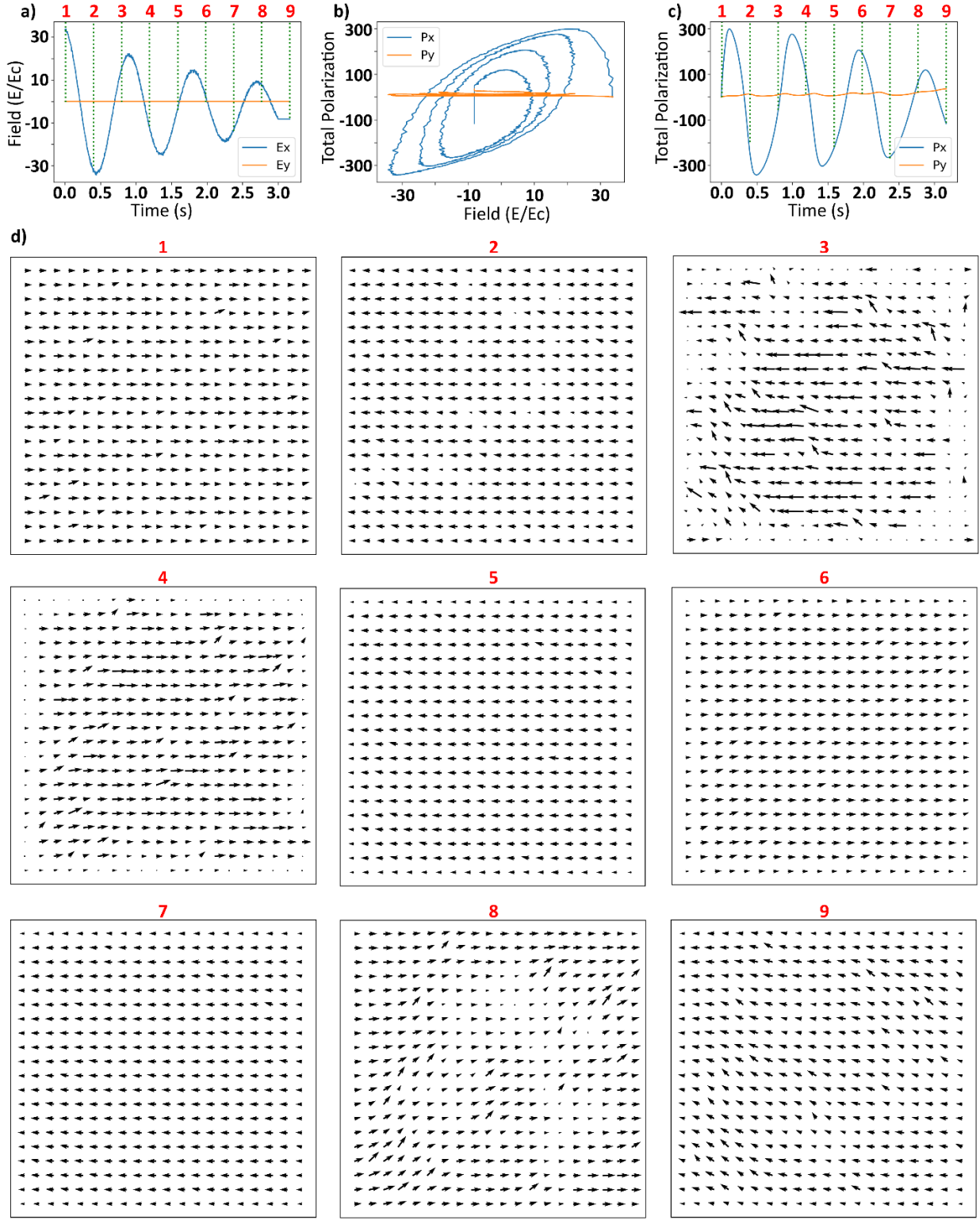

**Figure 4.** (a) Applied external electric field as a proportion of coercive field of the system as a function of time, (b) Resulting polarization in x-direction and y-direction as a function of applied electric field in x-direction (c) Resulting polarizations in x and y directions as a function of time,

(d) The snapshots of polarization vectors at each lattice sites at uniform time intervals. Vertical green lines in (a) and (c) represent the time at which polarization maps in (d) are plotted.

The results of an example simulation of FerroSIM are shown in shown in Fig. 4. Here the FerroSIM is run on a 20*20 lattice with impurities placed at 15% of these sites. These impurities are placed at randomly selected locations can create local field disturbances in either x or y directions. For simplicity, external electric field applied is applied in only x-direction and is shown in Fig. 4a. The electric field applied is a sinusoid with a fixed frequency whose amplitude is modulated with a positive exponential function. This external electric field is applied for 3 seconds divided into 900 times steps. The system is then equilibrated for a span of 50-timesteps at a constant field strength of electric field at $t=3s$. The resulting polarization in x and y components as a function of the applied field are shown in Fig. 4b and as a function of time is shown in Fig. 4c. Polarization in x-direction as a function of time has a similar shape as the applied electric field in x-direction and is lagging which is typical to a kinetic simulation of the ferroelectric materials. Polarization in y-direction oscillates with a very low amplitude due to impurities causing local fields in y-direction and the coupling between the polarizations in x and y directions. Polarization as a function of the applied electric field in x-direction encompasses the properties associated with the traditional ferroelectric materials like hysteresis loop, coercive fields, remnant polarization. Finally, polarization plots at different time steps are shown in Fig. 4d. These time steps are also marked with green lines in Fig. 4a and Fig. 4c. At lower applied electric field strengths, the local electric field is dominated by the impurities. Random field impurities do not allow the polarization to completely align with the external electric field at lower field strengths. The coupling mechanism between the lattice sites enforced in the FerroSIM propagates this behavior to the nearby neighbors. This results in formation of domains whose polarization is different from the overall polarization of the lattice. This phenomenon can be observed in the $8^{th}$ snapshot of Fig. 4d. This results in a non-zero curl of the polarization. On the contrary at higher electric field strengths, polarization vectors of all the lattice sites are aligned with the external electric field which would result in zero curl. The final curl of the lattice also depends on the history of the electric field. If the electric field in the recent past is large compared to the fields due to impurities, all the effects due to the impurities will be negated.

With the FerroSIM model illustrated, we proceed to apply the proposed optimization method for the process optimization, specifically the history of the electric field. As an initial family of possible trajectories, we chose family of sinusoids whose amplitudes are modulated by exponentials. These set of trajectories are then optimized to find the trajectory that results in maximum sum of absolute curl magnitude at each lattice site. To realize this on a 20*20 lattice at a given concentration of the impurities, we have created a set of electric field trajectories in the form of,

| | $A exp(\alpha t) \sin(\omega t) + B$ | (9) |
|---|---|---|

where $A \in [0, 0.75]$, $\alpha \in [-2.75, 2.75]$, $\omega \in [-2.75, 2.75]$. The parameters for each curve are randomly sampled from uniform distributions with the above-mentioned limits. The exponential factors are considered in such a way that the curve still has some visible sinusoidal structure whereas the frequency of the sinusoids is considered so that the electric filed has 3-4 cycles in the time considered. The curves are then generated for 900 equidistant timesteps corresponding to 3s using equation. 9 and then set to a constant value which is the final value of the field at the 900$^{th}$ timestep for the next 50-timesteps. This is done so that the system equilibrates at a constant external field for fifty timesteps before the curls are calculated. Each of these curves are normalized to fall in [-1, 1]. Since the curves with positive exponential coefficients always have their maxima at the end i.e., at t=3s, after normalization, the final value of these curves at t=3s is always 1. This would mean that the half the curves generated equilibrate at the same electric field and might not result in a variation in curl. To negate this effect and to include the effect of the exponential factor on the final value of the electric field, we let curves develop for 1200 timesteps, normalized the curve, and then considered only the first 900 timesteps of the curve. The 50 equilibration time steps are then added later. We have performed this operation on the exponentially increasing and decreasing curves to be consistent.

These trajectories are then projected into a two-dimensional latent space using VAE as shown in Fig. 5. This smooth continuously varying latent space will be sampled to generate the electric field that belong to the family of curves described above. Twenty-five randomly drawn electric field trajectories from the family of electric field curves generated are shown in Fig. 5a. The two-dimensional latent space is uniformly sampled, decoded, and shown in Fig. 5b. The latent distributions i.e., trajectories plotted as points in the 2-D latent space and are colored as a function of $\alpha$ in Fig.5c, $\omega$ in Fig. 5d, and $B$ in Fig. 5e. The latent space appears to be a strong function of $\alpha$ and $\omega$, and they vary smoothly and continuously over the latent space. The latent space does not appear to be strongly correlated with $A$ or $B$ of the input curve. The effect of $A$ and $B$ on the latent space is lacking since each curve is normalized prior to the application of VAE.

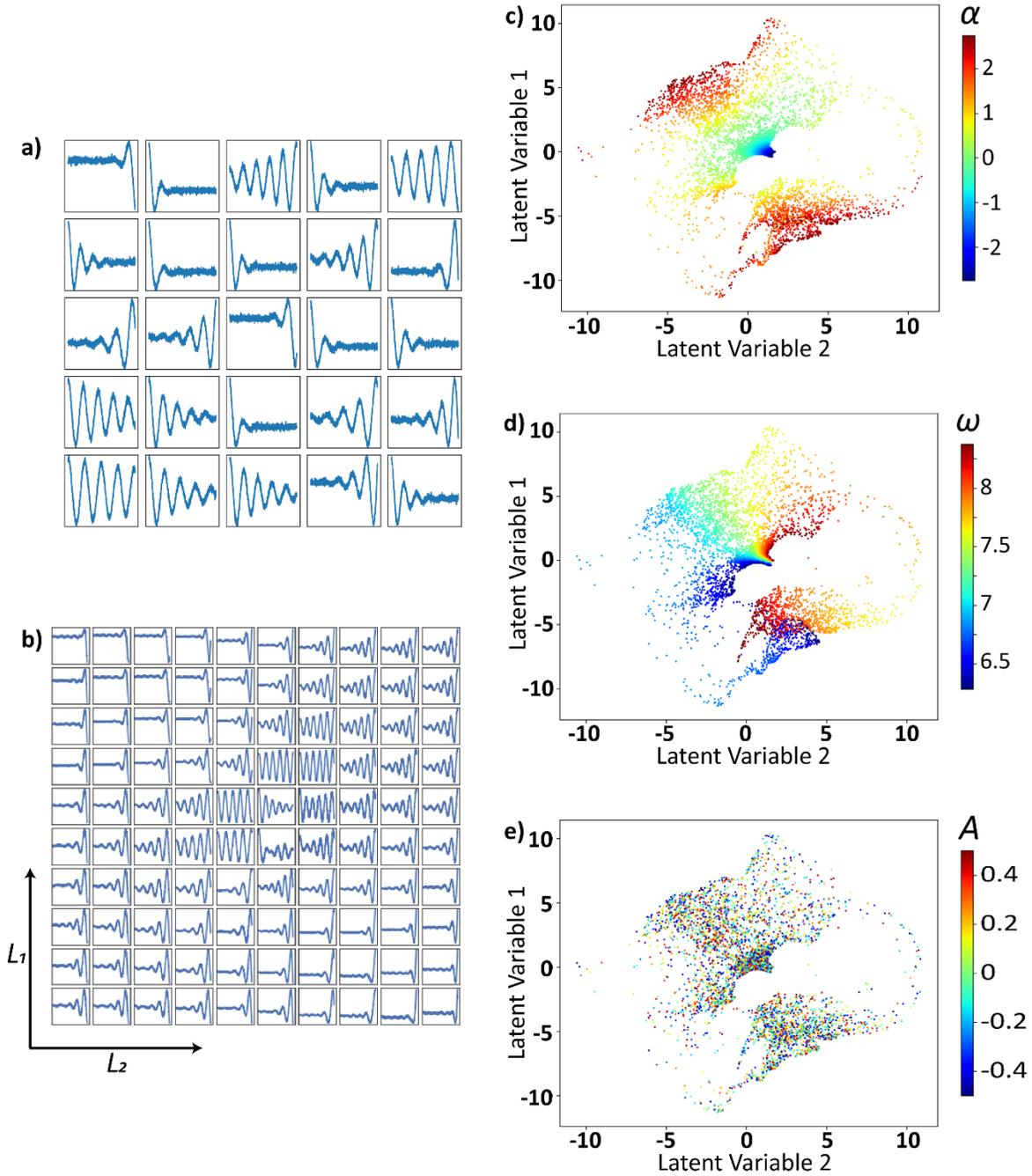

**Figure 5.** (a) Twenty-five randomly drawn curves from the input space of electric fields to the VAE, (b) Uniformly sampled and decoded representation of the latent space. The input functions are represented as points in the latent space and are colored using their respective values of (c) α, (d) ω, and (e) A

    The latent space is uniformly sampled into 10,000 points and then decoded back into the space of electric fields to study the variation of the sum of absolute curls. The sum of magnitude of the curl at each lattice site will be referred to as the curl of the system from hereon for brevity.

These decoded electric field curves are also normalized by design and are multiplied by a constant (150) to make them practical for the simulations. This constant value is user selected and the discussions that follow depend on the value of this multiplication constant. Once a curve is decoded, only first 900-timesteps of the curve is used as an input to the FerroSIM, discarding the last 300-timesteps as mentioned earlier. The equilibration timesteps, where the external electric field is held at a constant value for 50-timesteps, are then added to all the curves. The FerroSIM simulations are then run for each of these electric fields corresponding to the 10,000 points in the latent space. The value of curl at the end of 950-timesteps is calculated for each point in the latent space and is shown in the middle in Fig. 6.

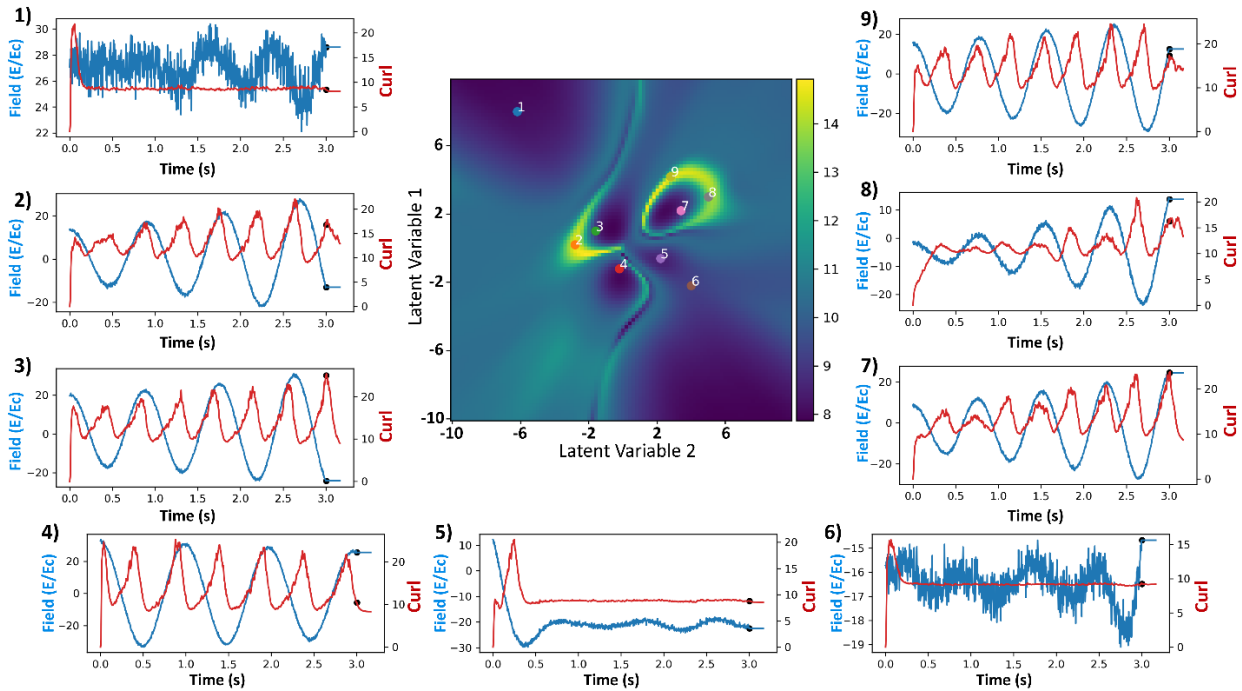

**Figure 6.** (Center) Sum of curl magnitudes at the end of the simulations as a function of latent space. Nine salient points are selected from the latent space and their respective histories of curl (red) and applied fields (blue) as a proportion of the coercive field are shown around the curl surface. The onset of equilibration region is marked as a scatter point in both curl and electric field histories.

The latent space distributions in Fig. 5c-d show that the encoding of the electric field curves is non-linear. Even with a linearly encoded latent space, the curl surface would be non-linear due to the abundance of parameters in the system that control the final value of curl. To obtain insights into the ground truth of the system, electric field and curl histories from different points in the latent space are plotted around the curl surface in Fig. 6. These points are assumed to represent their local neighborhood in the latent space. Some parts of the latent space are relatively straightforward to explain. For example, points 1 and 6 refer to the curves with a high absolute value of the electric field and low standard deviation. Electric fields with this behavior will lead to large magnitudes of polarization vectors as the system always experiences a high value of external electric field. Larger values of polarization vectors decrease the $F_{loc}$ in equation. 2 by making the $-E_{loc}.p$ term larger. At such large values of polarization fields, curl is not sustainable as any deviation from the average polarization vector will lead to a very high value of the coupling term in equation. 5 and thereby increasing the free energy of the system. This results in absence and/or minimization of the curl at these locations. The rest of the points in the latent space exhibit two common trends: i) the curl appears to be maximum at the local optima of the electric field, and ii) in the equilibration region, the curl continuously decreases in magnitude. The apparent overlap of the maxima in curl and the local optima in electric fields seems counterintuitive. Polarization, lagging the applied electric field, is about to reach its own optimum at the optima of the applied electric fields. Furthermore, near the maximum value of polarization, as explained above, any deviation from the average polarization field of the lattice is not sustainable. For the second trend, the rate of decrease of curl seems to be proportional to the value of curl at the onset of the equilibration region. For example, comparing points 2 and 3, the value of curl at the onset of the equilibration region is ~ 25 and 17, respectively. The curl in the equilibration region corresponding to point 2 falls rapidly to a low value. The curl corresponding to point 3 falls slowly and belongs to one of the two regions that produce a local maximum in the final curl surface.

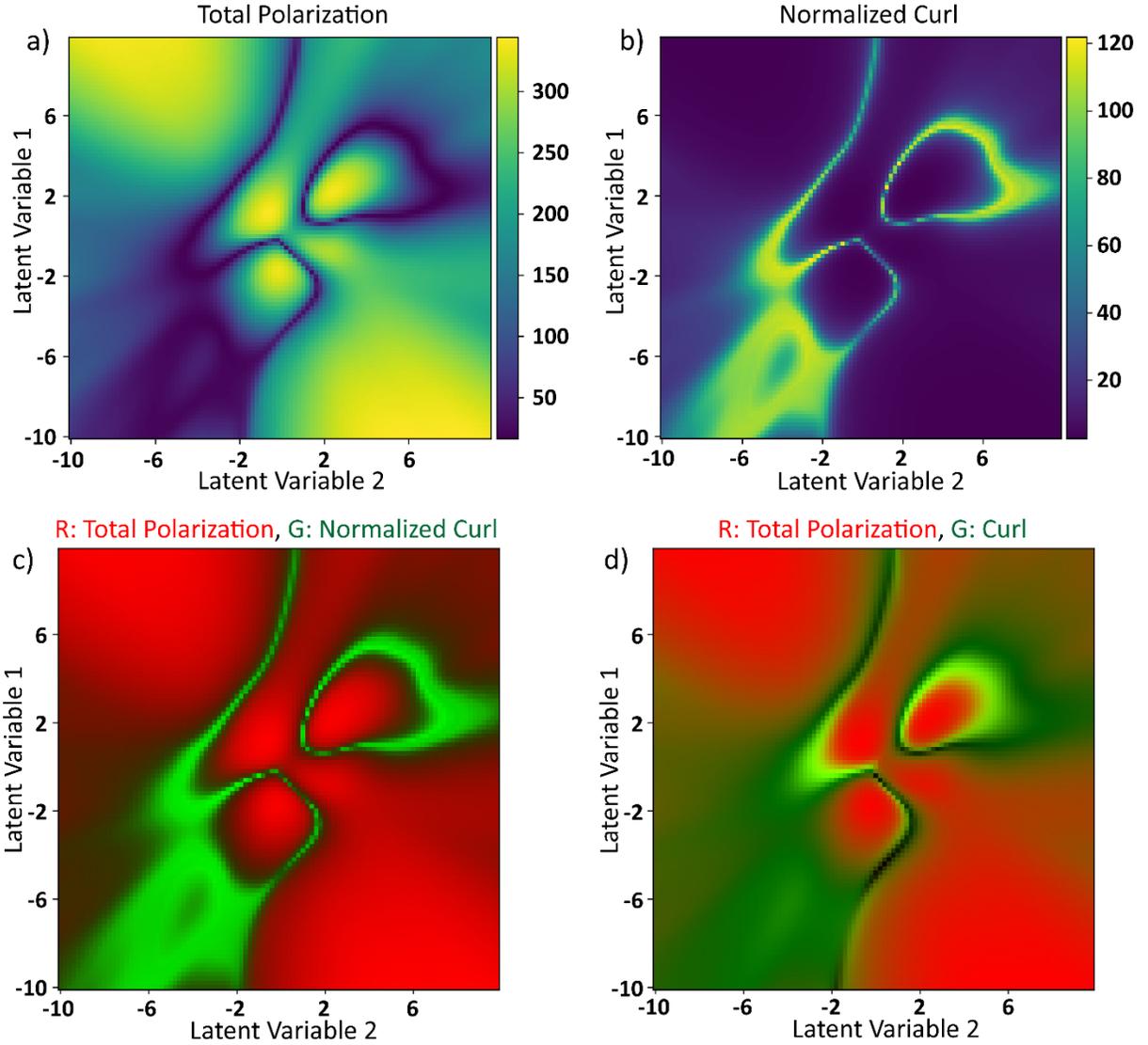

**Figure 7.** (a) Total polarization magnitude as a function of latent space, (b) Unit Curl: Sum of absolute curls as a function of latent space when the polarization vectors at each lattice site are normalized at the end of 950-timesteps, (c) Total polarization magnitude and Unit Curl as a function of latent space are represented as red and green channels in an RGB image, (d) Total Polarization magnitude and curl surface from Fig. 6 are represented as red and green channels respectively in an RGB image. Blue channel is set to be zeros in both (c) and (d).

To explore effects of the final value of total polarization on the resulting curl of the system, we plotted the total polarization at the end of the simulation as a function of latent space in Fig. 7a. To negate the effects of magnitude of polarization on the final value of curl, the curl is recalculated after normalizing the polarization vectors at each lattice site at the end of the simulations. The simulations are however run using the actual polarization vectors for the entire duration. The recalculated curl (referred to as normalized curl) is shown in Fig. 7b. The normalized curl surface will provide insights on how much the polarization field rotates without accounting for the magnitude of polarization vectors. It can be observed from Fig. 7a and Fig. 7b that the polarization surface and normalized curl surface are inversed with respect to each other i.e., the normalized curl is maximum when the polarization is minimum and vice versa. To visualize this better, the total polarization surface is used as the R-channel and the normalized curl is used as the G-channel while the B-channel is left empty in an RGB image and is shown in Fig. 7c. It can be observed that most part of the image is either entirely green (large values of normalized curl) or entirely red (large values of polarization) with very little overlap. This is because when the magnitude of the polarization is small, any curl (deviation from the average polarization field) in the system will not lead to a substantial coupling term in free energy equation (equation. 5). These deviations from the average polarization vector will lead to a large value of the normalized curl. However, since any such deviations at large polarization fields are not sustainable and any small deviations will not contribute much to the normalized curl, its value in these regions is going to be small. But the final value of curl is also dependent on the actual magnitude of the polarization vectors. A field with higher magnitudes of polarization vectors will result in higher value of curl compared to a field with smaller polarization vectors but a similar rotational tendency. To show this, the total polarization is used as the R-channel and the curl surface from Fig. 6 is used as the G-channel in an RGB image and is shown in Fig. 7d. It can be observed that there is some overlap between the green and red regions.  So far from the discussion, it is apparent that neither a large nor a small value of the total polarization corresponds to the maximum curl. The polarization needs to be at an optimal value to create a large curl.

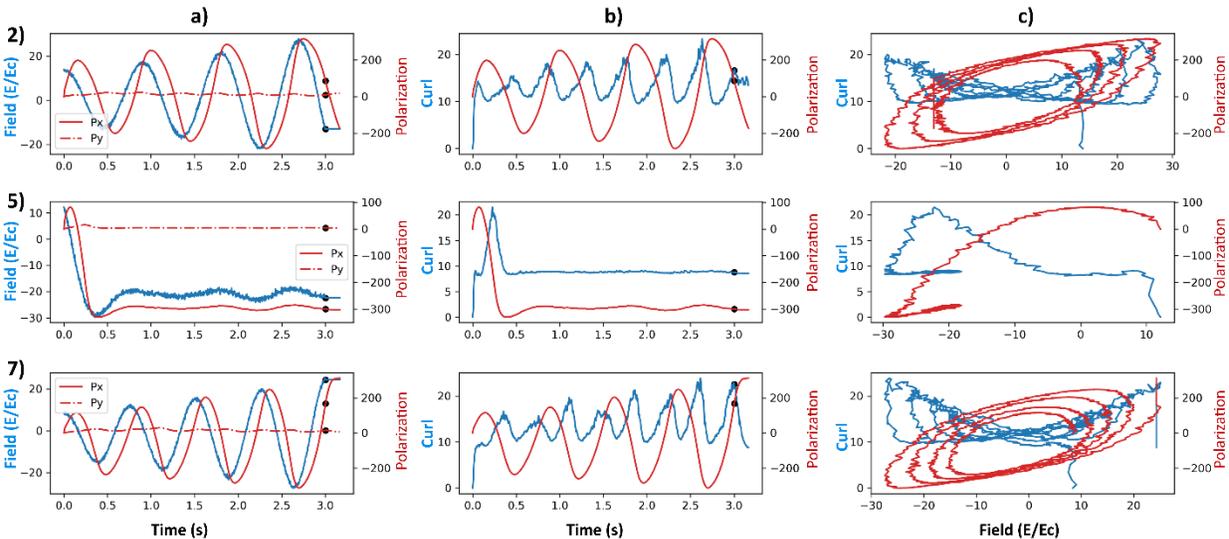

**Figure 8.** (a) Total polarization in x-direction (red), total polarization in y-direction (red-dotted), and curl (blue) as a function of time, (b) Curl (blue) and total polarization magnitude (red) as a function of time, (c) Curl (blue) and total polarization magnitude (red) as a function of applied electric field. The applied electric field is represented in the units of coercive field of the system.

The rows represent the points - 2, 5, 7 in the curl surface from Fig. 6. The onset of equilibration region is marked with a black scatter point in (a) and (b).

We carried out additional analyses to investigate the coincidence of local maxima of the curl surface and the local optima of electric field (Fig. 8). Points - 2, 5, and 7 from Fig. 6 are considered for this analysis. Point 2 corresponds to a large value of final curl, point - 7 corresponds to a large value of curl when the system starts to equilibrate which then falls rapidly to a very small value of curl. Finally, point-5 corresponds to a very large negative exponential factor ($\alpha$) in the electric field which causes the electric field to decay to a constant value in less than a cycle. For these three points in the latent space, polarizations in x and y directions and the applied electric field as a function of time are plotted in Fig. 8a. Curl and polarization magnitudes as a function of time are plotted in Fig. 8b, and as a function of electric field in Fig. 8c. It can be observed from Fig. 8a that the polarization lags the electric field in every case. This is intuitive as the effect always lags the cause in any kinetic simulation (akin to learning rate in ML). From Fig. 8b it can be observed that the curl is maximum after a few timesteps once the polarization crosses zero. This is because the polarization vectors are in a state of maximum normalized curl right after the coercive field. Subsequently, the polarization vectors start to grow in magnitude because of a high value in the applied electric field. It takes a few time steps in the simulation for the polarization to grow to a magnitude that constitute the maximum curl for the electric field's half cycle. In our simulation, the half cycle of electric field is approximately same as the time steps the curl takes to attain the local maximum. This results in apparent coincidence of the local optima of the electric fields with the local maxima of the curl which can also be observed in Fig. 8c. This local maximum of curl is not stable as the electric field has a very high absolute value at its optima and destroys the curl in the subsequent time steps. This destruction of curl is what results in a low final value of curl for point – 7. For point – 5, since the electric field is always a high value and so are the polarization fields, any small deviations in the polarization field might result in some curl. The results so far depend on the various parameters that go into the FerroSIM simulation. Any change in these parameters will lead to a different curl surface but the underlying physics will still be same.

Finally, we demonstrate the Bayesian optimization of processing trajectories in the latent space via the maximization of the curl. Note that the chosen material model generally is not prone to the natural formation of polarization vortices, and hence non-zero curl can emerge only due to the interactions between the frozen disorder and polarization field for a given field history. The curl distribution in the latent space is obtained by a brute force search where the simulations are run for every point in the latent space. When we are only concerned about optimizing the magnitude of the curl and/or when the latent space is dimensionality is large making the process of running simulations at each latent point intractable, one can utilize the Bayesian Optimization towards a guided search of the latent space.

Bayesian Optimization is an extension of Gaussian Processes (GP) regression. GP regression is an approach of interpolating or learning a function *f*, given the set of observations D = {($x_1$, $y_1$), . . .($x_N$ , $y_N$ )}. It is assumed that the arguments $x_i$ are known exactly, whereas the observations can have some Gaussian noise with zero mean in them i.e., $y_i = f(x_i) + \varepsilon$. The key assumption of the GP method is that the function *f* has a prior distribution f ~ $\mathcal{GP}(0,K_f(x, x'))$, where $K_f$ is a covariance function (kernel). The kernel function defines the relationship between the values of the function across the parameter space, and its functional form is postulated as a part of the fit. Here we implemented the Matern kernel function given by equation .10

$$k_{Matern}(x_1, x_2) = \sigma^2 \exp\left(-\sqrt{5} * \frac{|x_1 - x_2|}{l}\right)\left(1 + \sqrt{5} * \frac{|x_1 - x_2|}{l} + \frac{5}{3} * \frac{|x_1 - x_2|^2}{l}\right) \quad (10)$$

The learning is performed via Bayesian inference in a function space and the expected value of the function, corresponding uncertainties, and kernel hyperparameters are optimized simultaneously. The output of the GP process is then the predicted data set, uncertainty maps representing the quality of prediction, and kernel hyperparameters. The aspect that makes GP standout from the other extrapolation methods by providing not only the function value, but also the uncertainty associated with the prediction. Bayesian Optimization (BO) then samples the input space based on an inbuilt or user defined acquisition function which uses the predictions by GP. The acquisition function can be set to a pure exploration state where the point selected by BO is either random or proportional to the uncertainty of the prediction. It can also be set in an exploitation state where the selection of next point in the input space is based on the target value of the function. In our case, the BO is set to explore 40% of the time while selecting the points randomly and exploit the rest 60% trying the find the maximum of the curl in the latent space.

The curl surface obtained through brute force is a result of running 10,000 FerroSIM simulations. For BO based GP, we ran 15% of the simulations and the rest of the points are interpolated by GP. At each iteration, BO samples a point in the latent space based on exploration-exploitation trade off and the simulation corresponding to that point in the latent space is run and the actual value of curl is obtained. This point is now added to the input space to GP and the process is repeated for 100 iterations. The results of BO-GP analysis are shown in Fig. 9.

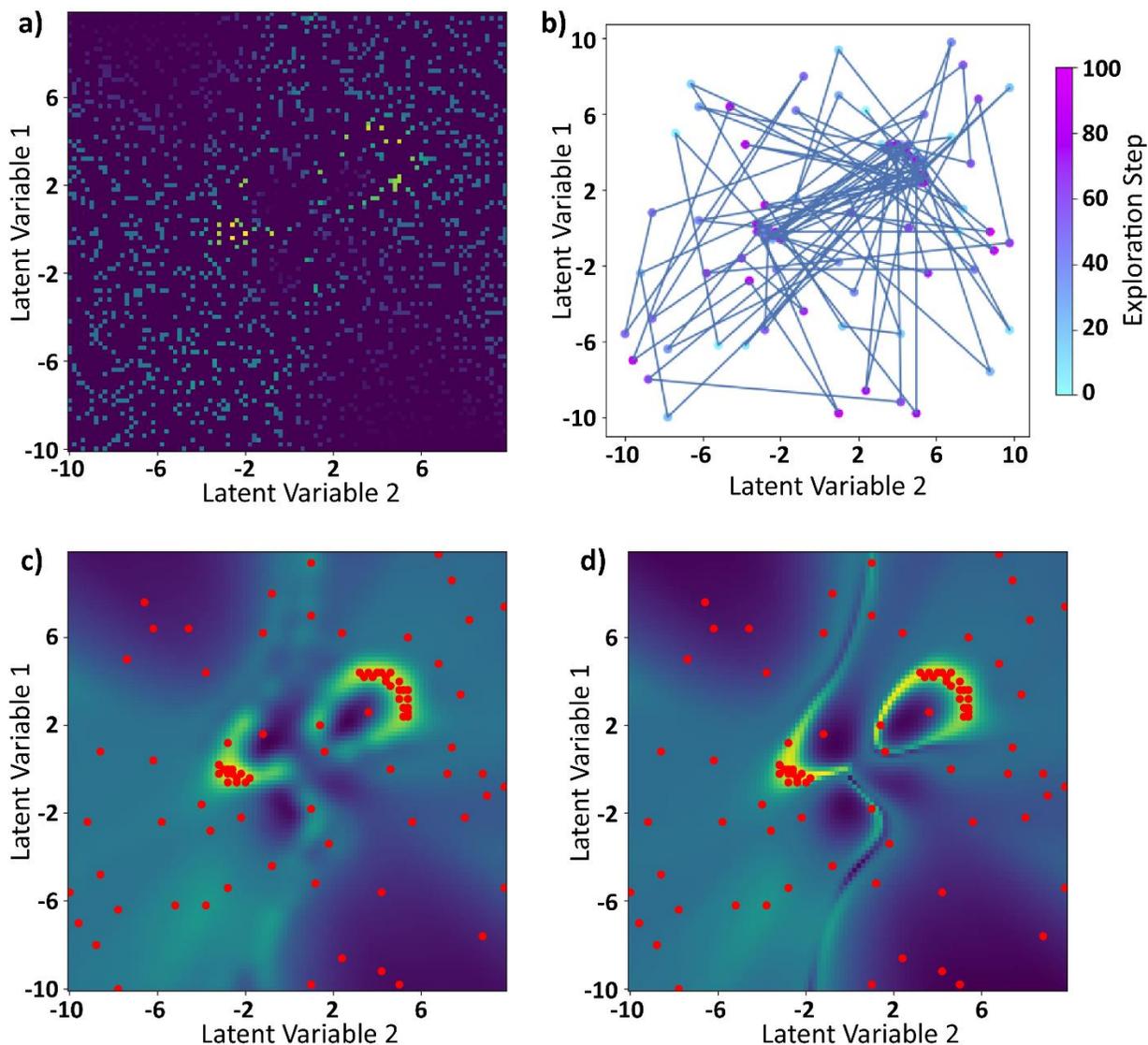

**Figure 9.** (a) Randomly sampled 15% of the total points (100*100) in the latent space that are provided initially as the seed points to GP-BO, (b) Points explored by BO in the latent space at the end of 100 iterations as function of the iteration number, (c) The curl surface approximated by BO at the end of 100 iterations, (d) Actual curl surface obtained through brute force calculations. The red scatter points in (c) and (d) are the points explored by BO in 100 iterations.

The seed points at the start of the BO process are shown in Fig. 9a. These are the 15% of the total points the latent space was divided into. Fig. 9b shows the point explored by the BO optimization in 100 iterations progressively. GP interpolations of the curl surface can be plotted in the Jupyter notebook attached in the data availability statement. Finally, the explored points are plotted in the GP predicted curl surface at the end of 100 iterations in Fig. 9c and the brute force curl surface in Fig. 9d. BO does a sub optimal job in exploring one of the two local maxima that is because the curl surface is multimodal with multiple local optima and none of the initial seed points to the BO are a part of this region. However, a mere 16% of the points are explored at the end of this exercise and the BO was able to identify the two local maxima of the curl surface even with incomplete information.

To summarize, here we proposed the approach for optimization of processing trajectories towards target functionalities via Bayesian optimization over the low-dimensional latent space of variational autoencoder. The initial training set here is formed based on the domain knowledge or human intuition. The effects of the variability of the initial trajectory set and distribution of parameters in the latent space are explored.

This approach is further used to maximize specific functionalities of material represented via lattice Ginzburg Landau model. Here we have demonstrated that this approach can be used to tune the process towards maximization of the total curl of polarization field in the system, which is a symmetry breaking phenomenon inconsistent with the symmetry of the lattice or applied stimulus. The maximization of the curl is found to occur in the vicinity of the phase transition as a result of the spontaneous symmetry breaking in the vicinity of the defects.

While here this approach is applied for specific family of the trajectories and model, we believe this approach to be universally applicable to a broad range of optimization problems such as materials processing, ferroelectric poling, and so on. It should be noted that the choice of the initial trajectory set has to be based on domain specific consideration and chosen broad enough to have sufficient variability but at the same time sufficiently narrow to avoid overly localized minima.

**Acknowledgements:**

This work was supported by the Energy Frontier Research Centers program: CSSAS–The Center for the Science of Synthesis Across Scales–under Award Number DE-SC0019288, located at the University of Washington (original idea and prototypes - S.V.K.), and the modeling and process optimization by the US Department of Energy Office of Science under the Materials Sciences and Engineering Division of the Basic Energy Sciences program (S.M.V. and R.K.V.). The Bayesian optimization research was supported by the Center for Nanophase Materials Sciences (CNMS) which is a US Department of Energy, Office of Science User Facility at Oak Ridge National Laboratory.

**Data Availability Statement:**

- The python code delineating the entire workflow of the paper is available at https://github.com/saimani5/Notebooks_for_papers/blob/main/FerroSim_VAE_BO_code_for_paper.ipynb The datasets required to reproduce the results can be downloaded inside the Jupyter Notebook.
- The code for FerroSim model can be found at https://github.com/ramav87/FerroSim/tree/rama_dev_updated

- The code for Variational Autoencoder can be found at https://github.com/ziatdinovmax/pyroVED
- The code for Gaussian Processes and Bayesian Optimization can be found at https://github.com/ziatdinovmax/GPim